\documentclass[10pt]{article}

\usepackage{inputenc}
\usepackage{graphicx}
\usepackage[rel]{overpic}
\usepackage{subfigure}
\usepackage{amssymb}
\usepackage{amsmath}
\usepackage{color}
\usepackage{hyperref}
\usepackage{soul}
\usepackage[numbers,super,sort&compress]{natbib}
\usepackage[bottom=2cm,top=2cm,left=2cm,right=2cm]{geometry}
\usepackage{rotating}
\usepackage{sectsty}
\usepackage{subfigure}
\usepackage{lettrine}
\sectionfont{\large}

\definecolor{new_blue}{rgb}{0.0,0.0,1.0}
\definecolor{orange}{rgb}{0.83,0.53,0.145}

\renewcommand{\eqref}[1]{equation (\ref{#1})}

\global\long\def\Rv{\mathcal{R}_v}

\begin{document}

\onecolumn
~
\vspace{3cm}

\vspace{3cm}

\noindent \textsf{{\huge 
Crackling Noise in Fractional Percolation -- Randomly distributed discontinuous jumps in {\em explosive} percolation}\\
~\\
\noindent Malte Schr\"oder$^{1,2}$, S. H. Ebrahimnazhad Rahbari$^{3}$, and Jan Nagler$^{1,2}$}

\vspace{2cm}

\noindent \textsf{ 
Crackling noise is a common feature in many systems that are pushed slowly,
the most familiar instance of which is the sound made by a sheet of paper when crumpled.
In percolation and regular aggregation clusters of any size merge until a giant component dominates the entire system.
Here we establish `fractional percolation' where the coalescence of clusters that substantially differ in size are systematically suppressed.
We identify and study percolation models that
exhibit multiple jumps in the order parameter where the position and magnitude of the jumps are randomly distributed  - characteristic of crackling noise.
This enables us to express crackling noise as a result of the simple concept of fractional percolation. In particular,
the framework allows us to link percolation with phenomena exhibiting non-self-averaging and power law fluctuations such as Barkhausen noise in ferromagnets.
}

\vspace{2cm}

\noindent \textsf{\textbf{Affiliations:}\\
$^1$ Max Planck Institute for Dynamics \& Self-Organization, 37077 G\"{o}ttingen, Germany\\
$^2$ Institute for Nonlinear Dynamics, Department of Physics, University of G\"{o}ttingen, 37077 G\"{o}ttingen, Germany\\
$^{3}$Department of Physics, Plasma and Condensed Matter Computational Laboratory,  Azarbayjan Shahid Madani University, Tabriz, Iran\\
~\\
\noindent \textbf{E-mail adresses:}\\
jan@nld.ds.mpg.de\\
}

\thispagestyle{empty}

~
\vspace{3cm}




\newpage

\begin{figure}[ht]
\centerline{
\includegraphics[width=8cm]{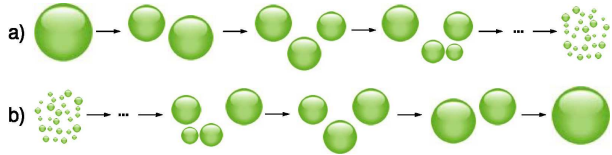}
}
\caption{\label{fragmentation}
{\bf Fragmentation and fractional percolation.}
a) In the process of fragmentation, clusters split up into parts of a certain fraction.
b) The reverse dynamics, `fractional percolation', is studied here.
{
In contrast to ordinary aggregation processes, in fractional percolation 
%
the coalescence of clusters that substantially differ in size  is systematically suppressed.
}
}
\end{figure}


\lettrine{\textcolor{orange}{M}}{any} 
systems crackle when pushed slowly.
Examples include the crumpling of paper \cite{paper}, earthquakes \cite{gutenberg}, solar flares \cite{solarflares},
the dynamics of superconductors \cite{superconductors}, and the magnetization of slowly magnetized magnets. 
For a piece of wood in fire one can even hear crackling noise without  special equipment.
Across all systems that display crackling noise,
the order parameter of the system exhibits randomly distributed 
{ jumps}, and
discrete, spontaneous events span a broad range of sizes \cite{sethna01}.
Magnification of the hysteresis curve of a magnetic material in a changing external field, for instance,
reveals that the magnetization curve is not smooth but exhibits small discontinuities. 
This series of correlated 
 {jumps} is called the Barkhausen effect, which is a standard example for crackling noise in physics \cite{magnetism, zapperi1, zapperi2}.
{ Despite its importance, crackling noise is far from being understood.
}

We study a simple random network percolation model  
that enables us to explain the key characteristics of crackling noise, 
within an analytical framework. To further demonstrate the universality of our approach,
computer simulations for the proposed percolation mechanism in geometrical confinement are carried out.

In random network percolation a fixed number of nodes are chosen randomly, and two of them are connected according to certain rules \cite{book1,book2, shlomo, spencer}.
This procedure is repeated over and over again until every {node} is connected to every other.
In the model studied here, connections between two clusters are preferred which are similar in size.
However, before the comparison is made 
the larger cluster is scaled down by a factor of $f$, referred to as the (possibly time dependent) target fraction.

The reverse dynamics of this percolation model represents simple fragmentation, see fig.\ \ref{fragmentation}.
Fragmentation processes where homogeneous parts break up into smaller ones are ubiquitous and have been studied intensely.
The applications range from atomic nuclei, and the fragmentation of glass rods, to fracture in large scale systems \cite{rods1, rods2, rods3, herrmann1990, sornette92b, krapivsky00, sornettebook}.
{
An important observation is that the size of the fragments are of the same order of magnitude as the parent pieces.
Thus the case where one fragment is microscopic while the size of the other fragment is substantially larger is rare.
We model this by systematically suppressing asymmetric break ups, to a degree controlled by the parameter $f$.
This suppression results in a preferential fractional increase of clusters that is time-reversed fragmentation.
}

Crackling noise and random network percolation are seemingly conceptually incompatible since the order parameter in percolation, the size of the largest component,
is believed to be globally continuous and would thus not fluctuate - except at the single point of the (first) phase transition \cite{book1, aharony96, riordan, riordan2012}.
In contrast, we demonstrate that crackling noise in percolation unexpectedly emerges from a simple fractional growth rule.

After all,
how can a process where links are successively added to a networked system account for crackling noise?
Which microscopic mechanisms imply stochastically distributed discontinuous transitions of the order parameter?
And finally, how robust are these properties?

To answer these questions, 
we study the nature of the fluctuations in the size of the largest component of a percolation model.
{
The particular model we use to exemplify the fractional growth mechanism can be replaced by any other model
where first a fixed number of nodes are chosen at random, and then two nodes are connected, 
according to any rule that forbids the largest chosen component to merge with components smaller than a fixed fraction 
of its size.
}
%

%
As we will reveal by a single event analysis, the network model features three basic properties:
(i) a fractional growth mechanism, (ii) a threshold mechanism, and (iii) a mechanism that amplifies critical fluctuations.
We show that these underlying mechanisms account for the main features 
of crackling noise. 
Perhaps most importantly, 
the framework allows us to derive macroscopic features from the underlying micro-dynamical mechanisms, 
which exposes connections between the seemingly unrelated concepts of percolation, fragmentation, and crackling noise.

\begin{figure}[ht]
\centerline{
\includegraphics[width=8cm]{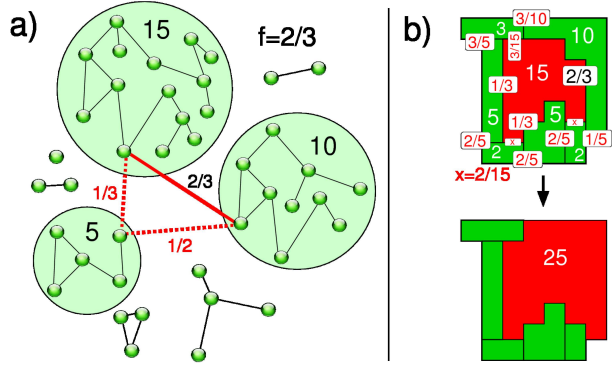}}
\caption{\label{model}
{\bf Sketch of the models.}
a) Select three nodes at random and calculate the sizes of the clusters they reside in $S_1 \ge S_2 \ge S_3$.
Connect 
those two nodes $v_i$, and $v_j$, which reside in clusters that minimize $\Delta_{ij}:=f S_i-S_j$ among $\Delta_{12}, \Delta_{13},$ and $\Delta_{23}$,
where $0<f\le 1$ is fixed.
(Actually, it is unnecessary to consider $\Delta_{13}$ since $\Delta_{12} \le \Delta_{13}$).
Here a link between the clusters of size 15 and 10 is established because
the (arbitrarily chosen) fixed target fraction $f=2/3$ is here exactly met:
$\Delta_{12}=2/3\times15-10=0 < \Delta_{23}=2/3 \times 10-5=5/3 <  \Delta_{13}=2/3 \times 15 - 5=5$.
b) Lattice model. Draw randomly a focal cluster (red) and merge those two neighbor clusters that minimize $\Delta$ (see text).
}
\end{figure}

\newpage
\section*{Results}

\paragraph{Network Model}

\begin{figure}[ht]
\centerline{
\includegraphics[width=8cm]{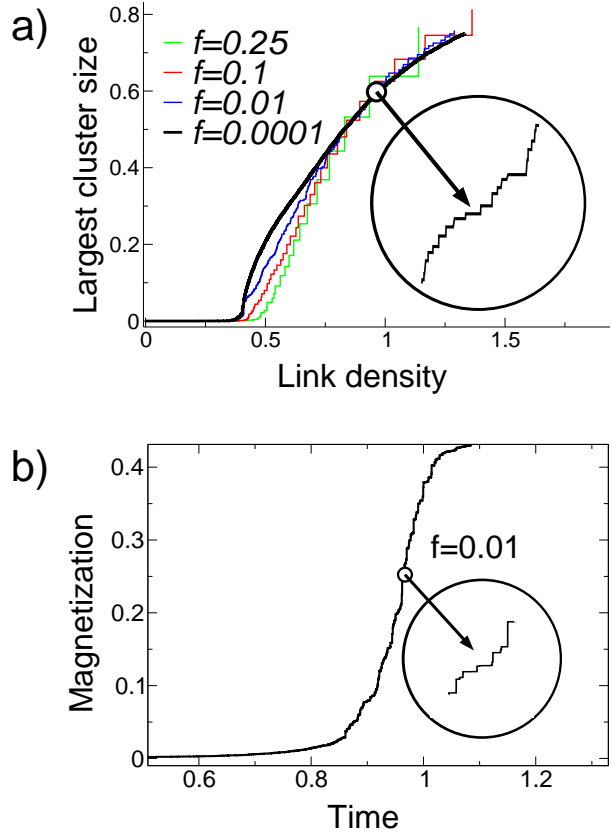}
}
\caption{\label{barkhauseneffect}
{\bf Discontinuous jumps.}
a) Scheme of crackling noise in network percolation.
Single realizations of fractional percolation processes for the network size $N=2^{18}$.
The evolution of the largest component $S_1/N$ is displayed.
The control parameter $f$ determines the magnitude of the discontinuities.
In the thermodynamic limit $N\rightarrow\infty$, the process shows infinitely many discontinuous jumps.
%
%
b) Same for the magnetization of the Barkhausen model on a square lattice, $N=400\times 400$, single realization for $f=0.01$. 
}
\end{figure}

Consider a network with a fixed number of nodes $N$ and $L$ links.
Start with $N$ isolated nodes and no links, $L=0$.
At each step, choose three different nodes $v_1$, $v_2$, and $v_3$ uniformly at random. 
Let $S_1$, $S_2$, and $S_3$ denote the sizes of the (not necessarily distinct) clusters they reside in.
Assume $S_1 \ge S_2 \ge S_3$, and $0<f\le 1$ is fixed.
Connect those two nodes $v_i$ and $v_j$ for which $\Delta_{ij}:=f S_i-S_j$, $1\le i < j \le 3$ is minimal, see fig.\ \ref{model}.
If necessary, choose randomly among multiple minima, whose corresponding nodes are to be linked.
Hence, a certain type of size homophily among clusters is applied where
connections between two clusters are preferred which are similar in size, after the size of the larger cluster has been rescaled by a factor $f$, the `target fraction'.
Since $\Delta_{12} \le \Delta_{13}$ 
only $\Delta_{12}$ and $\Delta_{23}$ have to be considered.  
%
The rule is also applied if the nodes to be linked reside in the same component.
As a ``final rule'', when there are only two clusters left in the system, connect these.
For single realizations of the process, see fig.\ \ref{barkhauseneffect}. 
As we will see, fragmentation as $S \rightarrow (gS:=S_i,(1-g)S:=S_j)$ with $g:=\frac{1+f}{1+2f}$,  is the inverse process 
{
and
the target fraction $f$ determines the magnitude of the discontinuities in the order parameter.

Let us order the largest components of the system by $\mathbf{S_1},\mathbf{S_2},\hdots$, with sizes $s_1=S_1/N \ge s_2=S_2/N \hdots$,
and write `with high probability' (whp) to express that the probability of a certain statement gets arbitrarily close to 1 as $N\rightarrow\infty$.
The link density of the network is the {analogue} of the occupation probability for lattices and defined by $p=L/N$ where $L$ denotes the number of links that have been added to the network. Let $p_c$ characterize the critical link density, the position of the (first) phase transition.
}

{
We show next that an arbitrarily fractional increase of components features 
discontinuities that survive even in the thermodynamic limit, and that this implies non-self-averaging \cite{sornettebook}.
}

{
\paragraph{Self-averaging}

A thermodynamic quantity, such as the total magnetization or the size of the largest component in a networked system $s_1$,
is self-averaging  if its relative variance becomes zero in the thermodynamic limit \cite{sornettebook},

\begin{equation}
\Rv:=\frac{\langle s_1^2\rangle-\langle s_1 \rangle^2}{\langle s_1\rangle^2}\rightarrow 0,
 \;\; \text{for} 
\; N\rightarrow \infty,
\end{equation}
where the brackets denote ensemble averaging.

For non-self-averaging systems, however, the thermodynamic quantity remains broadly distributed for large systems and large sample sizes.
Systems that lack self-averaging therefore lack the collapse of the ensemble average, and its minimum and maximum as well.
Non-self-averaging plays an important role in the statistical physics of disordered systems, 
for instance in spin glasses \cite{spinglasses1, spinglasses2, sornettebook}, neural networks, polymers, and population biology \cite{derrida81,parasi88}.

We characterize non-self-averaging in percolation by a non-vanishing relative variance of the order parameter $s_1$ on an extended interval.
For investigating this it is helpful to study the underlying microscopic mechanisms in terms of a single event analysis.

}

\paragraph{ Fractional growth mechanism}

First we show that, for $f>0$, the largest component cannot merge with components smaller than $\frac{f}{1+f}S_1$.
If less than three distinct clusters are picked, either an intracluster link is added and the size of largest component is necessarily unchanged,
or the size of the largest component doubles.
Thus we consider the case of three distinctly chosen components whose sizes are ordered, $S_1'\ge S_2'\ge S_3'$.

Proof by contradiction:
Assume $\Delta_{1,2}$ is minimal such that $\mathbf{S_1'}$ and $\mathbf{S_2'}$ merge when (A) $S_2' < \frac{f}{1+f} S_1'$.
In fact, $\Delta_{1,3}$ is never minimal (except if it is equal to another $\Delta$) since $\Delta_{1,3}\ge \Delta_{1,2}$, and if $\Delta_{2,3}$ was minimal then the largest chosen cluster would not merge with any other.
{
By multiplying A with $f$ and adding $-S_3'$ we obtain $f S_2'-S_3' < \frac{f}{1+f} f S_1' -S_3'$.
Since $\Delta_{1,2}$ is minimal, we have $f S_1' - S_2' < f S_2'-S_3'$,
and hence we obtain
$f S_2'-S_3' < \frac{f}{1+f} ( (1+f)S_2'-S_3') - S_3' = f S_2' - (1+\frac{f}{1+f}) S_3'$
which is impossible for $f>0$.
}

{
Thus, $S_1'$ either stays constant, increases `fractionally' by at least a factor of $\frac{f}{1+f}$}, or is overtaken (by a merger of $\mathbf{S_2'}$ and $\mathbf{S_3'}$).
However, overtaking becomes unlikely as the size of the largest component increases.

\paragraph{Impossibility of $\mathcal{O}(N)$ overtaking}

By $\mathcal{O}(N)$ overtaking, we mean the merger of two components, each smaller than the largest component,
which together are larger than the largest component and of size $\mathcal{O}(N)$.
Our line of arguments holds for any rule based on picking at most three nodes randomly. 
Assume that $S_1=\mathcal{O}(N)$, considering the following cases.

Case (i): Both smaller components are $\mathcal{O}(N)$.
This is (whp) impossible because the upper limit for the number of macroscopic components is 2.
Actually for any {$n$-node rule (where first $n$ nodes} are chosen randomly followed by any other rule),
there cannot exist more than $n-1$ macroscopic clusters over any extended period of time \cite{nagler2012}.

Case (ii): At least one of the smaller components is $o(N)$.
In this case, overtaking is (whp) impossible since either $S_1 \ge S_2' + S_3'$, or $S_1 \rightarrow S_1 + o(N)$.


\paragraph{Threshold mechanism}

Taken together, for the infinite system,
this implies that as soon as 
the size of the second largest component exceeds $\frac{f}{1+f}s_1$,
the second largest component merges with the largest one, $s_1 \rightarrow s_1 + s_2$.
Since a third macroscopic component is (whp) impossible, this also implies the reset of the second largest component, $s_2 \rightarrow 0$.

%

\paragraph{Power law fluctuations by amplification of critical fluctuations}

Summarizing the above considerations, after the first phase transition, for $p>p_c$ and $N\rightarrow \infty$,
the size of the largest component either stays constant or jumps discontinuously.
Since the first transition is point-continuous \cite{riordan, riordanX}
the process  necessarily exhibits infinitely many discontinuous transitions arbitrarily close to the first transition point, $p=p_c$.

\begin{figure}[ht]
\centerline{
\includegraphics[width=10cm]{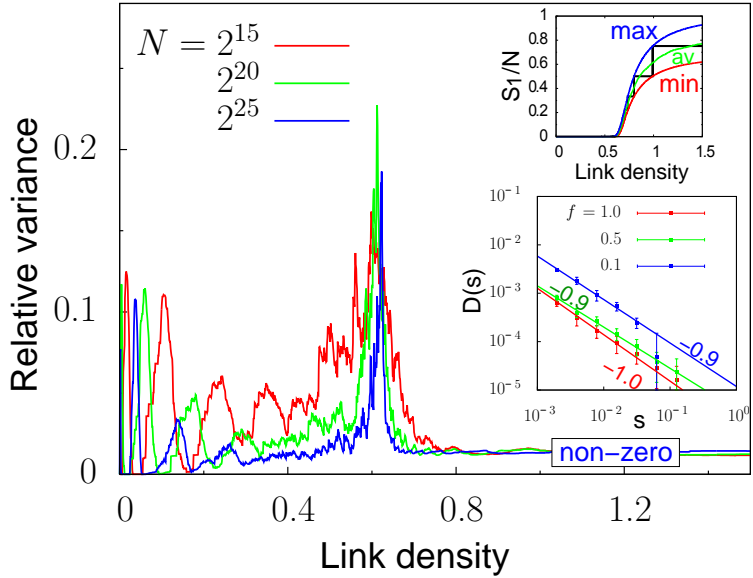}
}
\caption{\label{nSA}
{
{\bf Non-self-averaging crackling noise.}
The relative variance $\Rv$ of the largest component  in dependence on the link density $p$ is displayed, $f=1$.
For $p>p_c$ the system is non-self-averaging characterized by $\Rv\rightarrow \text{const.}>0$, for $N\rightarrow\infty$.
Upper inset: Stochasticity of transition points characterized by the lack of the collapse of average, minimum and maximum values of $S_1$, 
here shown for $f=1$ and an ensemble of 500 realizations.
A single realization is displayed in black.
Lower Inset: As derived in the text, for any $f>0$, $N\rightarrow\infty$, the jump sizes are power law distributed, $D(s)\sim s^{-\tau}$ ($\tau=1$ theory).
Fit exponents $\tau_{f=1}=0.96 \pm 0.05$ ($R=0.990$), $\tau_{f=0.5}=0.85 \pm 0.02$ ($R=0.998$), and
$\tau_{f=0.1}=0.90 \pm 0.04$ ($R=0.997$).
Error bars indicate standard deviations.
%
%
}
}
\end{figure}

Let $\delta_n$ denote the height of the $n$th step down the staircase.
{
The fractional growth mechanism suggests
the proportionality $\delta_n \sim g^n$ where $g=(1+\frac{f}{1+f})^{-1}$.
}
This yields the jump size distribution 
\begin{equation}\label{Ds}
D(s)\approx \int \delta(s-\delta_n)\text{d}n = \int \delta(s-\gamma g^n)\text{d}n \sim {s}^{-1},
\end{equation}
for $f>0$, where $\delta(\cdot)$ denotes the Delta function and $\gamma$ is a constant.
This is supported by numerics (see fig.\ \ref{nSA}).

The stochasticity is a consequence of the exponential amplification of the critical fluctuations  of $s_1$ at the first phase transition point, $p_c$.
{ Fluctuations, measured by the relative variance, in the size of the largest component $s_1$ at $p_c$ are known to be non-zero, $\Rv(s_1(p_c))> 0$ \cite{aharony96, book1}. 
}

From $\delta_n \approx \varepsilon g^n$ we see that uncertainties $\varepsilon$ are exponentially suppressed as $n$ increases.
In contrast, fluctuations in $s_1$ at $p_c$ are exponentially amplified, for increasing $p$.
As a result, both the size of the jumps, and the transition points are stochastic - even for the infinite system.

Thus, the process is non-self-averaging, characterized by a non-vanishing relative variance 
\begin{equation}
\Rv:=\frac{\langle s_1^2\rangle-\langle s_1 \rangle^2}{\langle s_1\rangle^2}>0,
 \;\; \text{for} \; p\ge p_c, \; N\rightarrow \infty.
\end{equation}
This is numerically supported, see fig.\ \ref{nSA}, and stands in contrast to the weakly discontinuous case (see Supplementary Notes 1-3).

Expressing $\delta_n$ in terms of a time dependent target fraction $f(n)$, 
and assuming $f(n)=\alpha/n$, $0<\alpha\le1$, from 
eq.\ (\ref{Ds}) we obtain
$D(s)\sim s^{-\frac{1+\alpha}{\alpha}}$ characterizing
power law fluctuations that decay faster than $1/s$.
Thus, other fluctuation types than $1/s$ are accessible via a non-constant $f$. 

\begin{figure}[ht]
\centerline{
\includegraphics[width=8cm]{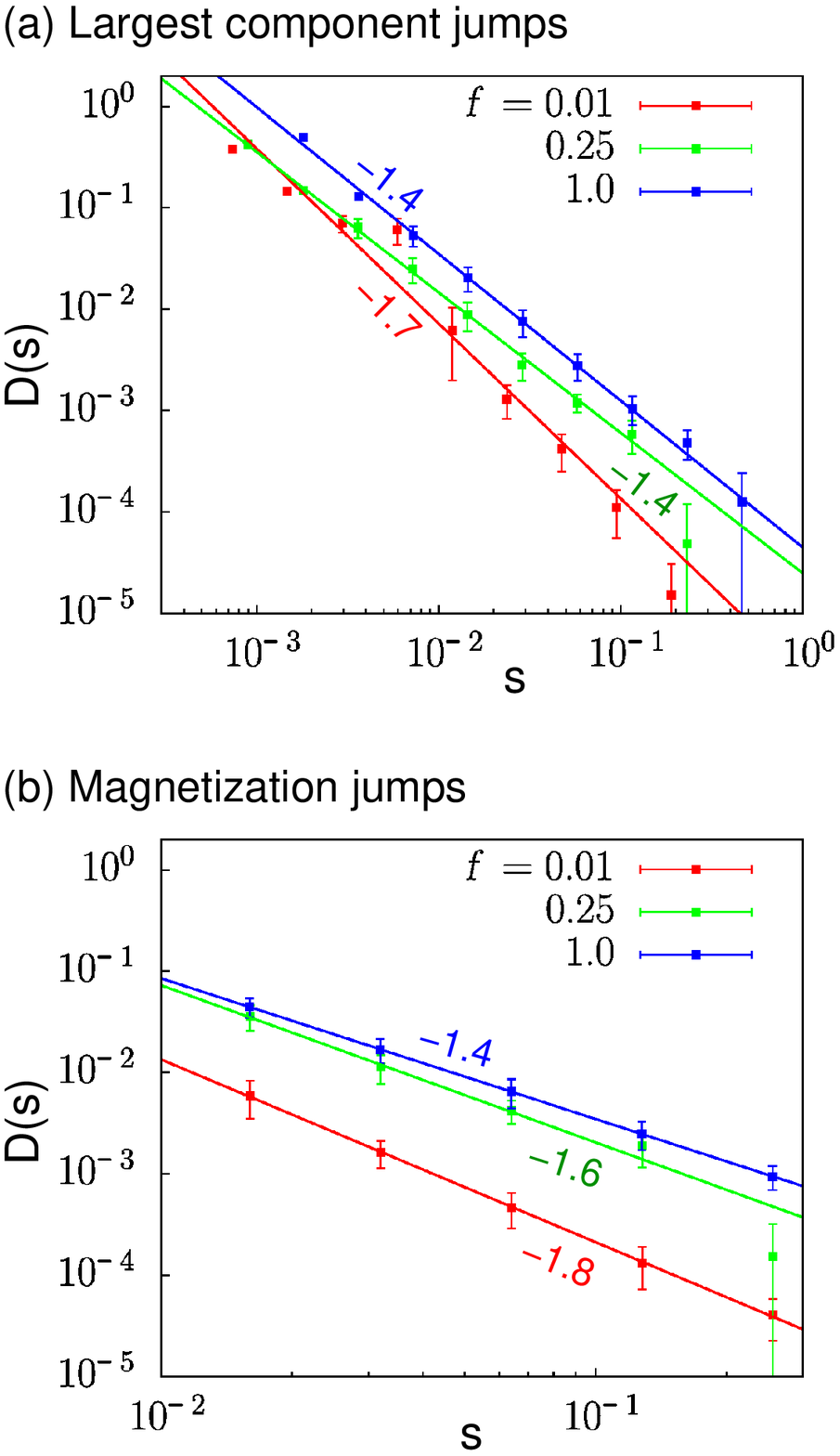}
}
\caption{\label{pdfs}
{
{\bf Power law fluctuations.}
Power law fluctuations (a) in the largest cluster size, and (b) in the total magnetization.
The jump size distributions $D(s)\sim s^{-\tau}$ suggest a power law decay. 
Lattice size $N=500\times 500$ (periodic boundary conditions). Initial condition with a single magnetized cluster.
Jumps smaller than the relative size $7\cdot10^{-4}$ are discarded.
Data points are averaged over 1000 realizations, using logarithmic binning. 
Distributions without using log-bins are shown in Supplementary Figure S6.
Error bars indicate standard deviations.
%
(a) Fit exponents $\tau_{f=0.01}=1.7 \pm 0.1$ ($R=0.983$), $\tau_{f=0.25}=1.4 \pm 0.03$ ($R=0.999$), and
$\tau_{f=1}=1.4 \pm 0.03$ ($R=0.999$).
(b) Fit exponents $\tau_{f=0.01}=1.8 \pm 0.02$ ($R=0.999$), $\tau_{f=0.25}=1.6 \pm 0.07$ ($R=1.00$), and
$\tau_{f=1}=1.5 \pm 0.03$ ($R=1.00$).
}
%
%
%
%
%
%
%
%
%
}
\end{figure}

\paragraph{Barkhausen percolation model}

Since clusters in our framework neither have a magnetization nor are geometrically confined, the analogy to magnetic effects such as the Barkhausen effect,
a standard example of crackling noise in geometrical confinement, remains incomplete.
In order to demonstrate the universality of our claims, next we study a  Barkhausen percolation model on a square lattice.
Assume that each cluster has a homogeneous magnetization; either $m(\mathbf{S}_i)=0$ or $m(\mathbf{S}_i)=1$.
Initially all sites are single clusters and have $m=0$, up to a set of sites of $o(N)$ that is set up to $m=1$, e.g., a few single clusters.
Now apply repeatedly the following update rule.
(i) Uniformly draw a cluster at random, (ii) among this focal cluster and its (von Neumann-)neighbors, merge those two neighbor clusters that minimize $\Delta:=f S_i-S_j$ (see fig.\ \ref{model}(b)),
(iii) magnetization: apply the neutral rules $0+0\rightarrow 0$, and $1+1\rightarrow 1$, together with the magnetization rule $0+1\rightarrow 1$, for the merging clusters.
In addition, apply the physical time increment rule $t\rightarrow t + \min(s_i,s_j)^{\frac{1}{2}}$ at each merger, $s_i$ and $s_j$ being the scaled cluster sizes of two merging components.
This rule accounts for cracks preceding a fragmentation.
Cracks have a finite propagation velocity, which implies that duration of a fragmentation event depends on the size of the fragments \cite{sornette92, sornettebook}.
Here, we have arbitrarily chosen the square root of the smaller cluster as the time increment.
However, the main features of the model are independent of the specific choice.

Fragmentation mimics the {repeated} reconfiguration of homogeneous magnetic domains under a slowly increased opposite external magnetic field.
A magnetic domain is a region within a ferromagnetic material with uniform magnetization.
During the demagnetization, domains split up into smaller ones of different magnetization, a process called reconfiguration.
This is in most ferromagnetic materials the dominating factor in the minimization of the local magnetostatic energy and accounts for the sudden jumps of the total magnetization in the hysteresis curve.
However, due to other effects, the process stops when the domain size approaches a threshold, usually in range of $10^{-4}$ to $10^{-6}$ m \cite{magnetism, sethna01}.

Here we demonstrate that the reverse process, fractional percolation, reproduces the main features of Barkhausen noise.
While only a caricature of the intricate processes in ferromagnets \cite{zapperi1, zapperi2}, it nonetheless explains multiple 
randomly distributed discontinuous jumps in the total magnetization $M(t):=\frac{1}{N}\sum S_i m(\mathbf{S}_i)$, 
together with non-SA, and power law fluctuations, see Fig.\ \ref{pdfs} and Supplementary Note 4.

\section*{Discussion}

We have established crackling noise in percolation.
In particular, we have demonstrated analytically that 
fractional growth rules 
imply randomly distributed jumps in the order parameter.
These jumps are discontinuous phase transitions.
However, 
when such mechanisms are mixed, even weakly, with 
mechanisms that merge components purely at random 
then the transitions vanish, or become at most weakly discontinuous
characterized by very small power law exponents \cite{nagler2011}, see Supplementary Note 2.

Fractional percolation describes nucleation where domains cannot grow in arbitrarily small pace.
As an application
consider an unmagnetized ferromagnetic sample of linear dimension of about $l=1\, cm$ at room temperature.
It is not unrealistic to assume that the magnetic domains have roughly the same linear dimension $l_0=10^{-3}\, cm$, independent of $l$, 
but different magnetizations that globally compensate each other. 
An increasing external magnetic field typically causes magnetic domains to increase at least by the size of one of its neighbor domains.
Thus this mechanism only would result in the total magnetization to either stay constant or jump in steps of
the size of magnetic domains.
As long as $l_0>0$ this quantized growth is an example of the fractional growth rule since the largest domain cannot increase by arbitrarily small amounts.
However, as we increase the sample size, the ratio $l_0/l$ decreases such that for the infinite system
any jump size becomes zero relative to the system size.
Thus our framework suggests that Barkhausen noise is at most `weakly discontinuous'.
In fact,
in many soft magnetic materials 
  Barkhausen jump sizes are not extensive and thus their relative size shrinks with increasing system size \cite{magnetism}.
In contrast, 
in thin magnetic films and other geometries where long range interactions are not of major importance 
macroscopic jumps have been reported \cite{berger2001}.

The characteristics of fractional percolation are robust against an arbitrary (time-dependent) variation of the parameter $f>0$ that determines the magnitude of the discontinuities.
The framework combines mechanisms reminiscent of many physical and biological systems:
the order parameter exhibits a sudden jump upon exceeding a dynamical threshold \cite{sethna01,btw1, btw2, sornette92}, and
large-scale fluctuations emerge as a consequence of critical fluctuations.
The amplification and propagation of critical fluctuations to macroscopic scales has been subject of intensive investigations in quantum critical systems,
such as the inflationary expansion of the early universe \cite{peebles80}, and disordered systems exhibiting quenched disorder \cite{spinglasses1, spinglasses2}.
However, the current understanding of most systems where randomness is frozen or amplified is far from being complete.
A recent study on group formation in small growing populations, for instance, shows that the fraction of one trait within the population (e.g., cooperators)
can be subject to strong fluctuations as a result of the amplification of stochastic fluctuations generated during the initial phase of the dynamics \cite{frey2010, frey2012}.

Power law fluctuations across operating scales, discontinuous jumps of the order parameter, and non-self-averaging may 
considerably subvert predictability and control of networked systems \cite{barabasi, shlomo}.
Exact conditions for these phenomena are elusive \cite{sornettebook}.
Our analysis provides sufficient conditions for these features.
Because the framework connects the seemingly unrelated concepts of percolation, fragmentation, and crackling noise,
it might help to qualitatively improve the understanding of systems that display (stochastic) discontinuous phase transitions.\\
In short, we expressed the main features of crackling noise as
 a consequence of a simple concept: fractional percolation.

\section*{Acknowledgments}
We cordially thank E. Vives, P. Grassberger, M. P. Touzel, and A. Trabesinger for valuable comments.
S.H.E.R.\ is supported by grant No.\ 90004064 from INSF.

\section*{Author contributions}
J.N. conceived the idea.
M.S. and J.N. contributed in designing the research.
M.S., J.N. and H.E.R. analyzed the data. 
M.S. performed the network simulations.  
H.E.R. performed the lattice simulations. 
M.S. and J.N. wrote the manuscript.

\section*{Competing financial interests}
The authors declare no competing financial interests.

\end{document}